\documentclass{pasj00}
\SetRunningHead{Astronomical Society of Japan}{Usage of \texttt{pasj00.cls}}

\usepackage{times}

\def\p1{\phantom{0}}
\def\x1{\phantom{00}}
\def\y1{\phantom{000}}

\begin{document}

\title{Application of the Titius--Bode Rule to the 55~Cancri System: \\
Tentative Prediction of a Possibly Habitable Planet
}
\author{Manfred Cuntz
}
\affil{Department of Physics, University of Texas at Arlington, Box 19059,\\
       Arlington, TX 76019, USA}
\email{cuntz@uta.edu}

\maketitle

\begin{abstract}
Following the notion that the Titius--Bode rule (TBR) may also be applicable
to some extra-solar planetary systems, although this number could be
relatively small, it is applied to 55~Cancri, which is a G-type main-sequence
star currently known to host five planets.  Following a concise computational
process, we tentatively identify four new hypothetical planetary positions
given as 0.081, 0.41, 1.51 and 2.95~AU from the star.  The likelihood that
these positions are occupied by real existing planets is significantly
enhanced for the positions of 1.51 and 2.95~AU in the view of previous
simulations on planet formation and planetary orbital stability.
For example, \citet{ray08} argued that additional planets would be
possible between 55~Cnc~f and 55~Cnc~d, which would include planets
situated at 1.51 and 2.95 AU.  If two additional planets are assumed
to exist between 55~Cnc~f and 55~Cnc~d, the deduced domains of stability
would be given as 1.3--1.6 and 2.2--3.3~AU.
The possible planet near 1.5~AU appears to be located at the outskirts
of the stellar habitable zone, which is however notably affected by
the stellar parameters as well as the adopted model of circumstellar
habitability.  We also compute the distance of the next possible
outer planet in the 55~Cnc system, which if existing is predicted
to be located between 10.9 and 12.2~AU, which is consistent with
orbital stability constraints.  The inherent statistical significance
of the TBR is evaluated following the method by \citet{lyn03}.  Yet
it is up to future planetary search missions to verify or falsify
the applicability of the TBR to the 55~Cnc system, and to attain
information on additional planets, if existing.
\end{abstract}

\begin{table*}
\caption{Stellar Parameters\label{SParm}}
\begin{center}
\begin{tabular}{lll}
\hline
\noalign{\smallskip}
Quantity                 & Value                                           & Reference \\
\noalign{\smallskip}
\hline
\noalign{\smallskip}
 Spectral Type           & G8~V                                            & \citet{gonz98} \\
 Distance$^a$            & $12.5 \pm 0.13$~pc                              & \citet{esa97}  \\
 Apparent Magnitude V    & 5.96                                            & SIMBAD website$^b$ \\
 RA Coordinate           & 08$^{\rm h}$~52$^{\rm m}$~35.811$^{\rm s}$      & SIMBAD website$^b$ \\
 DEC Coordinate          & +28$^\circ$~19$^\prime$~50.95$^{\prime\prime}$  & SIMBAD website$^b$ \\
 Effective Temperature   & 5280~K                                          & see text      \\
 Radius$^c$              & $0.925 \pm 0.023~R_{\solar}$                    & \citet{rib03}  \\
 Mass                    & $0.92 \pm 0.05~M_{\solar}$                      & \citet{val05}  \\
 Age$^d$                 & $5 \pm 3$~Gyr                                   & \citet{fis08}  \\
 Metallicity [Fe/H]      & $+0.31 \pm 0.04$                                & \citet{val05}  \\
\noalign{\smallskip}
\hline
\end{tabular}
\end{center}
\begin{list}{}{}
\item[]
$^a$Data from the {\em Hipparcos Catalogue}. \\
$^b$See {\tt http://simbad.u-strasbg.fr} \\
$^c$Alternative value: 1.15~$\pm$~0.035~$R_{\solar}$ \citep{bai08}. \\
$^d$A more stringent determination, which is $4.5 \pm 1$~Gyr, has been given
by \citet{don98} and \citet{hen00}.
\end{list}
\end{table*}

\begin{table*}
\caption{Planetary System$^a$\label{PlaSys}}
\begin{center}
\begin{tabular}{lrlrlrlc}
\hline
\noalign{\smallskip}
 Planet   &  \multicolumn{2}{c}{Distance~(AU)} & \multicolumn{2}{c}{Period~(d)} &
             \multicolumn{2}{c}{$M~{\sin}i$~($M_{\rm J})$} & Discovery \\
\noalign{\smallskip}
\hline
\noalign{\smallskip}
 55~Cnc~e & 0.01583 & $\pm$ 0.00020   &    0.7365400 & $\pm$ 0.0000030 & 0.027 & $\pm$ 0.0011 & 2004 / 2011 \\
 55~Cnc~b & 0.115   & $\pm$ 0.0000011 &   14.65162   & $\pm$ 0.0007    & 0.824 & $\pm$ 0.007  & 1996 \\
 55~Cnc~c & 0.240   & $\pm$ 0.000045  &   44.3446    & $\pm$ 0.007     & 0.169 & $\pm$ 0.008  & 2002 \\
 55~Cnc~f & 0.781   & $\pm$ 0.007     &  260.00      & $\pm$ 1.1       & 0.144 & $\pm$ 0.04   & 2007 \\
 55~Cnc~d & 5.77    & $\pm$ 0.11      & 5218         & $\pm$ 230       & 3.835 & $\pm$ 0.08   & 2002 \\
\noalign{\smallskip}
\hline
\end{tabular}
\end{center}
\begin{list}{}{}
\item[]$^a$Data given by \citet{fis08} and references therein, except for 55~Cnc~e, for which the data
from \citet{daw10} and \citet{win11} were used.
\end{list}
\end{table*}

\section{Introduction}

One of the more controversial aspects in the study of the Solar System
as well as studies of selected extrasolar planetary systems is the
application of the Titius--Bode rule (TBR).  Historically,
the TBR was first derived and applied to the Solar System where it
played a significant role in the search for new planetary objects
(e.g., \cite{nie72}).
The discovery of Uranus by Herschel in 1781 and the largest object of
the Mars--Jupiter asteroid belt, Ceres, by Piazzi in 1801 appeared to
confirm the applicability and significance of the TBR.  The TBR
historically given by the formula
\begin{equation}
r_n = 0.4 + 0.3 \times 2^n ,~~~~n = - \infty, 0, 1, 2, 3, ...
\end{equation}
(in AU) is, however, entirely inapplicable to Neptune and to the former
Solar System planet Pluto\footnote{Pluto is mentioned merely for
historical reasons; it is often considered to highlight the
limitations of the TBR for the Solar System.  Pluto is in the 3:2
mean-motion resonance with Neptune and is thus not a dynamically
independent object.  Moreover, long-term integrations have uncovered
that its orbit is chaotic yet stable over billion-year timescales
\citep{mal93}.}, which is still often considered to be an
appropriate representative of the steadily increasing number of
Kuiper Belt Objects (e.g., \cite{jew01}).

A considerable amount of previous discussion revolves around the
interpretation of the TBR for the Solar System, including its
well-known insufficiency, see, e.g., \citet{gra94}, \citet{hay98},
\citet{lyn03}, and \citet{nes04} for previous studies largely based on
statistical analyses.  For example, \citet{lyn03} argued that it is
not possible to conclude unequivocally that laws (or rules) of
Titius--Bode type are, or are not, significant, a conclusion that
also appears to be consistent with the work by \citet{hay98}.
Moreover, the main conclusion of \citet{lyn03} is also shared by
\citet{nes04}.  \citet{nes04} also pointed out the curiosity that
if the Earth's distance is excluded from the sequence, then the
probability of the occurrence of that sequence by chance is
radically reduced from 95 to 100\% down to 3\%.

If no appropriate physical explanation for the TBR is identified,
it is typically argued that the TBR may be merely a rule of chance
(``numerology"), fuelled by the psychological tendency to identify
patterns where none exist \citep{new94}.  On the other hand, there
are previous detailed astrophysical and astrodynamical studies that
potentially point to a physical background of the TBR, at least for
the Solar System and (by implication) for a selected number of extrasolar
planetary systems as well.  For example, \citet{whi72} argued that
jet streams may develop in a rotating gaseous disk at discrete orbital
distances given by a geometric progression.  This result appears to be
broadly consistent with the work by \citet{sch84}, who pursued an analytic
analysis of the structure, stability, and form of marginally stable
axisymmetric perturbations of idealized rotating gas clouds.  His
study showed that the radial parts of the expansion solutions obtained
from the governing differential yield simple oscillating functions with
TBR-type features.  

A similar approach was undertaken by \citet{gra94}, who argued that
a Titius--Bode type law emerges automatically as a consequence of the
scale invariance and rotational symmetry of the protoplanetary disc,
and that such geometrical relationships are a generic characteristic
of a broad range of physical systems.  \citet{li95} investigated the
nonlinear development and evolution of density disturbances in nebular
disks on the basis of one of their previous studies.  Their analysis
showed that the perturbed density is unstable with respect to self-modulation,
leading to the formation of density field localization and collapse.
In the case of a self-similar collapse, the perturbed density was found
to increase with time and to form steady rings with very large but limited
amplitudes.  The distances of these rings were found to follow a
geometric progression akin to the TBR.  Also, in his summary about
the final stages of planetesimal accumulation, \citet{lis93}
pointed out that ``[t]he self-limiting nature of runaway growth
strongly implies that protoplanets form at regular intervals in
semimajor axis."  This result appears to entail geometrical spacing
of planets in planetary systems, potentially modified by their
long-term orbital dynamical evolution, in possible agreement with
(a modified version of) the TBR.

Another approach invoking statistical numerical experiments was
adopted by \citet{hay98}.  They chose to fit randomly selected artificial
planetary systems to Titius--Bode type laws considering a distance rule
inspired by the Hill stability of adjacent planets.  They did not identify
a heightened significance of the TBR, except that its meaning is that
stable planetary systems tend to be spaced in a regular manner.
However, they indicate that their method could be used to identify
potentially unstable planetary systems, especially if applied in
conjunction with long-term orbit integrations.
A further contribution to the interpretation of the TBR was made
by \citet{not96} and \citet{not97} based on an approach of scale
relativity and quantization of the Solar System.  It attempts to
describe the Solar System in terms of fractal trajectories governed
by a Schr\"odinger-like equation.  The physical background and
justification of this approach, however, remain highly uncertain.
A controversial but testable aspect of this theory is that it proposes
the existence of one or two small planets between the Sun and Mercury,
which to date have no support through  observations.  On the other hand,
these studies point to the general possibility of ``empty" orbits,
which may also be of interest to generalized applications of the TBR
as well.

In the following, we will apply a generalized version of the TBR, without
subscribing to a specific physical interpretation, to the planetary system
of 55~Cancri (= $\rho^1$~Cnc = HD~75732 = HIP~43587 = HR~3522; G8~V) (see
Table~1), which was historically
the fourth star other than the Sun after 51~Peg, 70~Vir,
and 47~UMa (not counting the pulsar PSR~1257+12 as well as other
controversial cases) that was identified hosting a planet.  A previous
effort to apply the TBR to 55~Cnc has been pursued by \citet{pov08},
which will also be discussed in the following.  Additionally, we will
compare our results to a modified version of the TBR given by
\citet{lyn03} previously applied to the Solar System.

55~Cnc is known to host five planets, named 55~Cnc~b to 55~Cnc~f,
which were discovered between 1996 and 2007 by \citet{but97}, \citet{mar02},
\citet{mca04}, and \citet{fis08}; note that all planets have been detected
using the radial velocity method.  The positions of the planets, according
to their original determinations, are given as 0.038, 0.115, 0.240, 0.781,
and 5.77~AU, respectively, and their masses ($M_p \sin i$) range from 0.034
to 3.835~$M_{\rm J}$ (see Table~2; it also gives information on the
respective uncertainties).

However, a notable controversy emerged about
the innermost planet 55~Cnc~e.  \citet{daw10} argued that the
position of 55~Cnc~e is given as 0.016 rather than 0.038~AU based
on an improved method that allowed to distinguish an alias from the true
planetary frequency in a more reliable manner.  This revised distance
was subsequently confirmed by \citet{win11} based on continuous photometric
monitoring with the MOST (Microvariability \& Oscillations of STars) space
telescope.  Moreover, there is an ongoing discussion about the principal
possibility of additional planets in the 55~Cnc system, particularly in the
gap between 0.8 and 5~AU.  This speculation is fuelled by detailed orbital
stability studies, including studies involving putative Earth-mass planets
\citep{blo03,riv07,ray08,smi09,ji09}.  If such a planet exists, it would also
possibly imply the presence of a potentially habitable planet if appropriate
conditions are met (e.g., \cite{lam09}).

The star 55~Cnc is a middle-aged main-sequence star with a mass of approximately
0.95 $M_{\solar}$ \citep{val05}, with a stellar effective temperature of about
5250~K (see Table~3).  From the {\it Hipparcos} parallax of $79.8 \pm 0.84$~mas
\citep{esa97}, a luminosity of $0.61 \pm 0.04$ $L_{\solar}$
(see Table~1) can be deduced, which is consistent with its spectral type G8~V
\citep{gonz98}; see also discussion by \citet{mar02}.  The age of the star can
be obtained from the strength of the chromospheric Ca~II H+K emission, indicating
an age of $4.5 \pm 1$~Gyr \citep{don98,hen00}; see also \citet{bal97} for
previous work.  Another element of our work is the assessment of
55~Cnc's circumstellar habitability, which will be pursued following the
approach by \citet{kas93} and \citet{und03}; additionally, it will consider the
possible extension of the outer boundary of habitability as discussed by,
e.g., \citet{for97}, \citet{mis00}, and \citet{hal09}.

Our paper is structured as follows.  After the introduction of historical
aspects of the TBR and 55~Cnc itself, as done, we describe our methods and
results.  Emphasis will be placed on the calculational process concerning
the TBR based on four different measures of difference; they are applied
to assess the deviations between the positions of the five known planets
of 55~Cnc (i.e., 55~Cnc~b to 55~Cnc~f) and the predictions of the TBR,
thus allowing to constrain the unknown TBR parameters.  In addition, we
comment on the possibility of a habitable planet in the 55~Cnc system as
implied by the TBR.  Thereafter, we consider other relevant studies,
particularly studies regarding constraints from orbital stability simulations
and planet formation.
We also comment on a previous application of the TBR to the 55~Cnc system
given by \citet{pov08}.  Moreover, we provide an analysis of the statistical
significance for the TBR forwarded by this work, as well as of the previous
TBR version by \citet{pov08}, following the method of \citet{lyn03}.
Finally, we present our summary and conclusions.


\section{Methods and Results}

\subsection{Computation of the TBR Parameters}

The centerpiece of our study is the application of the TBR to the planetary system
of 55~Cnc.  Akin to the application of the TBR to the Solar System, we originally
selected as approach
\begin{equation}
a_i^{\rm TB} \ = \ \cases{A                    & \enskip if \enskip $i = 1$ \cr
                          A + B \cdot Z^{i-2}  & \enskip if \enskip $i > 1$ }
\end{equation}
guided by the original planetary data of \citet{fis08}.
Here $A$, $B$, and $Z$ constitute free parameters, which need to be determined
based on the known positions of the previously discovered five planets.  Based on
the five known planets 55~Cnc~e, 55~Cnc~b, 55~Cnc~c, 55~Cnc~f, and 55~Cnc~d, are
identified as $i=1$, 3, 4, 6, and 9, respectively.

This version of the TBR is consistent with its original version customarily
applied to the Solar System.  Note that $B$ indicates the compactness of the
system (i.e., the higher $B$, the less dense the overall planetary distribution),
whereas $Z$ indicates the progression of the relative spacing of the planets.
For the Solar System, the free parameters of the above given equation
take the following values: $A=0.4$, $B=0.3$, and $Z=2.0$.  However,
in the meantime, a controversy emerged regarding the location of the
innermost planet 55~Cnc~e, which based on a re-analysis of existing data
was identified to be positioned at 0.016~AU rather than 0.038~AU.  We
therefore revised the proposed version of the TBR such that the position
of the innermost planet is changed to $a_1^{\rm TB} = A/Z$.  Fortunately,
test simulations have shown that the key predictions of this work, i.e.,
the location of the hypothetical planets at 0.081, 0.41, 1.51 and 2.95~AU
as well as the possible exterior planet, expected to be located between
10.9 and 12.2~AU (if existing), are entirely unchanged by this alteration.
Moreover, the innermost planet is also not part of the polynomial fit
determined by $Z$, just like in case of the Solar System.

\begin{figure}
\begin{center}
\FigureFile(96mm,60mm){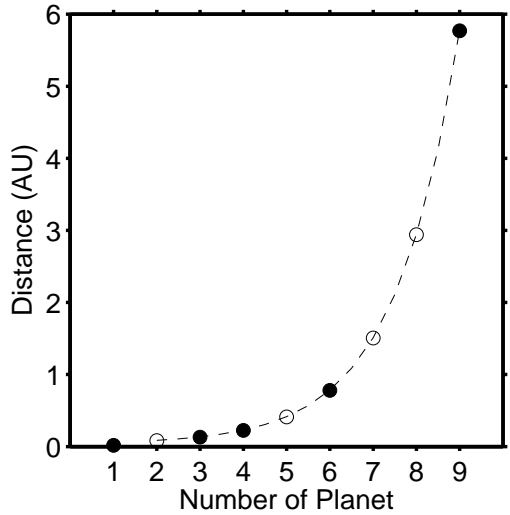}
\end{center}
\caption{
Distances of observed (full circles) and predicted (open circles)
planets (or planetary positions) in the 55~Cnc system.  The
predicted planets have been computed following the TBR based
on MDIF~1 (see Table~\ref{PlaSys}).
}\label{fig1}
\end{figure}

\begin{table}
\caption{Stellar Effective Temperature\label{STemp}}
\begin{center}
\begin{tabular}{ll}
\hline
\noalign{\smallskip}
$T_{\rm eff}$~(K) & Reference \\
\noalign{\smallskip}
\hline
\noalign{\smallskip}
  5150 $\pm$  75  &  \citet{gonz98}  \\
  5336 $\pm$  90  &  \citet{fuh98}   \\
  5250 $\pm$  70  &  \citet{gonv98}  \\
  5243 $\pm$  93  &  \citet{str01}   \\
  5338 $\pm$  53  &  \citet{rib03}   \\
  5279 $\pm$  62  &  \citet{san04}   \\
  5234 $\pm$  44  &  \citet{fis05}   \\
\noalign{\smallskip}
\hline
\end{tabular}
\end{center}
\end{table}

\begin{table*}
\caption{Titius--Bode Parameters\label{TBPar}}
\begin{center}
\begin{tabular}{lccc}
\hline
\noalign{\smallskip}
MDIF    &  $A$ & $B$ & $Z$ \\
\noalign{\smallskip}
\hline
\noalign{\smallskip}
MDIF~1  & 0.032 $\pm$ 0.003 & 0.049 $\pm$ 0.001 & 1.973 $\pm$ 0.006 \\
MDIF~2  & 0.031 $\pm$ 0.001 & 0.050 $\pm$ 0.001 & 1.971 $\pm$ 0.006 \\
MDIF~3  & 0.031 $\pm$ 0.002 & 0.050 $\pm$ 0.001 & 1.969 $\pm$ 0.006 \\
MDIF~4  & 0.031 $\pm$ 0.001 & 0.047 $\pm$ 0.001 & 1.991 $\pm$ 0.004 \\
\noalign{\smallskip}
\hline
\end{tabular}
\end{center}
\end{table*}

\begin{table*}
\caption{Distances of Real and Hypothetical Planets$^a$\label{PosRHP2}}
\begin{center}
\begin{tabular}{lccccc}
\hline
\noalign{\smallskip}
 Planet      &  Observed &  \multicolumn{4}{c}{Titius--Bode Rule} \\
\noalign{\smallskip}
\hline
\noalign{\smallskip}
           &           & MDIF~1 & MDIF~2 & MDIF~3 & MDIF~4 \\
\noalign{\smallskip}
\hline
\noalign{\smallskip}
 55~Cnc~e    &  0.016  &  0.016  &  0.016  &  0.016  &  0.015  \\
 ... (HPL~1) &  ...    &  0.082  &  0.081  &  0.081  &  0.078  \\
 55~Cnc~b    &  0.115  &  0.130  &  0.129  &  0.130  &  0.125  \\
 55~Cnc~c    &  0.240  &  0.224  &  0.224  &  0.225  &  0.218  \\
 ... (HPL~2) &  ...    &  0.411  &  0.411  &  0.413  &  0.403  \\
 55~Cnc~f    &  0.781  &  0.779  &  0.781  &  0.783  &  0.771  \\
 ... (HPL~3) &  ...    &  1.506  &  1.509  &  1.511  &  1.504  \\
 ... (HPL~4) &  ...    &  2.940  &  2.945  &  2.945  &  2.965  \\
 55~Cnc~d    &  5.77   &  5.770  &  5.774  &  5.768  &  5.872  \\
\noalign{\smallskip}
\hline
\end{tabular}
\end{center}
\begin{list}{}{}
\item[]
$^a$Distances in units of AU.
\end{list}
\end{table*}

The pivotal aspect of our study is to determine and optimize the parameters
$A$, $B$, and $Z$ based on the positions of the five known planets
of the 55~Cnc system, which are identified as $i=$ 1, 3, 4, 6, and 9. 
The positions of these five planets are calculated using adequate trial
values for $A$, $B$, and $Z$.  These parameters are subsequently optimized
through minimizing the numerical differences between the observed and
``predicted" distances (i.e., semimajor axes) of the observed five planets.
The later also requires a measure of difference.  In the following four
different measures of difference, henceforth referred to as MDIF, have
been adopted, which are: absolute linear deviation (MDIF~1), relative
linear deviation (MDIF~2),  absolute quadratic deviation (MDIF~3), and
relative quadratic deviation (MDIF~4).  The equations for the distance
difference parameter ${\delta}r_i$ are thus given as
\begin{equation}
{\delta}r_i \ = \  \Bigl\vert r_{pl,i} - r_{pl,i}^{\rm TB} \Bigr\vert
\end{equation}
\begin{equation}
{\delta}r_i \ = \  \Biggl\vert{{r_{pl,i} - r_{pl,i}^{\rm TB}} \over {r_{pl,i}}}\Biggr\vert
\end{equation}
\begin{equation}
{\delta}r_i \ = \  \Bigl( r_{pl,i} - r_{pl,i}^{\rm TB} \Bigr)^2
\end{equation}
\begin{equation}
{\delta}r_i \ = \  \Biggl({{r_{pl,i} - r_{pl,i}^{\rm TB}} \over {r_{pl,i}}}\Biggr)^2 \ ,
\end{equation}
where $r_{pl,i}$ denotes the observed distance (semi-major axis) of planet $i$
and $r_{pl,i}^{\rm TB}$ denotes the distance value obtained
through the application of the TBR.

Table~4 gives detailed information on the TBR parameters $A$, $B$, and
$Z$, including the respective uncertainty bars, which are due to the
distance uncertainties of 55~Cnc~b to f (see Table~1) given by
observations.  It is found that $A \simeq 0.031$, $B \simeq 0.049$, and
$Z \simeq 1.98$.  There is very little impact by the particular
choice of measure of difference.  For example, the spacing parameter
$Z$ is found to vary between $1.969 \pm 0.006$ (MDIF~3) and
$1.991 \pm 0.004$ (MDIF~4).  Furthermore, the distance
uncertainties for 55~Cnc~b to f, given by the observations,
have also a relatively small effect on the values of $A$, $B$,
and $Z$; note that the associated uncertainty is 0.3\% or less
in case of $Z$.

\begin{figure}
\begin{center}
\FigureFile(88mm,55mm){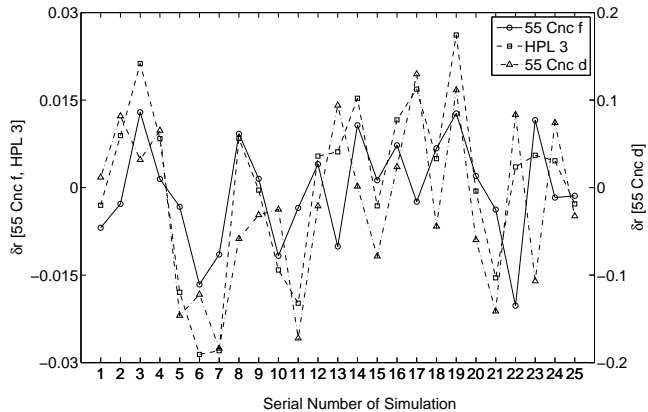}
\end{center}
\caption{
Distance variations ${\delta}r$ for 55~Cnc~f, 55~Cnc~d, and the
hypothetical planet 3 for a series of 25 statistical simulations.
For 55~Cnc~f and 55~Cnc~d, these variations are due to the observational
uncertainties, which are 0.007 and 0.11~AU, respectively.  For HPL~3,
the distance variations were computed following the TBR based on MDIF~1
(see Table~\ref{PlaSys}).
}\label{fig2}
\end{figure}

\subsection{Applications and Tests}

The derivation of the TBR parameters $A$, $B$, and $Z$ can be used
to identify possible additional planetary positions in the 55~Cnc
system (see Table~5), potentially occupied by planets.  Specifically,
four planetary positions are found, which are: 0.081, 0.41, 1.51 and
2.95~AU.  As expected from our previous discussion, the influence
of the different choices of MDIF is only of very minor importance.
Figure~1 offers a detailed depiction of the observed and predicted
hypothetical planets.  Note the very high quality of the fit
for $i > 1$.

Prior to discussing further results of our study, we want to
present the findings from detailed tests about the convergence
of the TBR parameters, notably $Z$.  Again, the TBR parameters
$A$, $B$, and $Z$ are obtained by trying to fit the positions
of the known planets 55~Cnc~b to 55~Cnc~f by using the TBR.
This requires the choice of a measure of difference, MDIF;
additionally, it also requires the choice of a precision parameter
for $A$, $B$, and $Z$ that was chosen as $10^{-k}$ with $k = 1$,
2, 3, and 4.  The control parameter of convergence was the
summed residual of the difference in the positions
between the observed and predicted (fitted) distances of the
five known 55~Cnc system planets.

Our results are given in Table~6.  As expected, it is found
that the higher the selected precision, the more accurate
are the obtained values for $A$, $B$ (not shown), and $Z$
deduced by fitting the positions of the five observed
planets.  However, the summed residuals are found to
never approach exactly zero, regardless of MDIF and the
precision parameter,
although the residual will converge toward a very small limit  (i.e.,
between $10^{-2}$ and $5 \times 10^{-4}$), especially if
MDIF~3 is selected.  This result is expected.
First, the higher the intended precision for $A$, $B$, and $Z$,
the smaller the resulting residual.  Second, the residual
cannot approach zero, even if an increasingly higher
precision for $A$, $B$, and $Z$ is permitted, owing to
the fact that the TBR (if applicable) should be
viewed as a ``rule" rather than a (physical) ``law".  There is
no way that the TBR as proposed (or any modification thereof)
will be able to precisely represent the positions of the
currently known five planets in the 55~Cnc system as also
encountered in applications of the TBR to the Solar System.

Next we focus on the case of the hypothetical planet HPL~3
(see Table~5) positioned near 1.51~AU; see Table~7 for
details.  Concerning the four measures of difference, MDIF,
the position of HPL~3 is found to range between $1.504 \pm 0.025$ (MDIF~4)
and $1.511 \pm 0.028$ (MDIF~3).
Note again that the uncertainty in the position of HPL~3
due to the choice of MDIF is insignificant.  However, this hypothetical planet
is of potentially high relevance because it may be located
in 55~Cnc's habitable zone (HZ).  Therefore, it will be
discussed in more detail below (see Sect.~2.3).

Considering the potential importance of HPL~3, if existing,
we also investigated the distance variation ${\delta}r$
due to the measurement uncertainties for 55~Cnc~f and 55~Cnc~d
(see Table~1) as published; the uncertainty measurement
is most substantial for 55~Cnc~d (both in absolute and
relative units).  The results are given in Fig.~2; here
the implemented version of the TBR was based on
MDIF~1.  Various positions for 55~Cnc~f and 55~Cnc~d
were assumed using a Monte--Carlo approach \citep{pre89}
considering their observationally deduced distance measurements. 
We pursued a total of 25 simulations.  In our series of
simulations, the absolute variation ${\delta}r$ for HPL~3
was less than 0.03~AU.

Another application of the TBR is the computation
of the distance of the next possible outer planet of the
55~Cnc system, i.e., located beyond 55~Cnc~d, which is
at a distance of approximately 5.77~AU.  This hypothetical 
planet is predicted to be located between 10.9 and 12.2~AU
(see Table~8).  More precisely, its position is predicted
in the range between $11.33 \pm 0.43$ (MDIF~3)
and $11.66 \pm 0.39$ (MDIF~4).

\begin{figure}
\begin{center}
\FigureFile(88mm,55mm){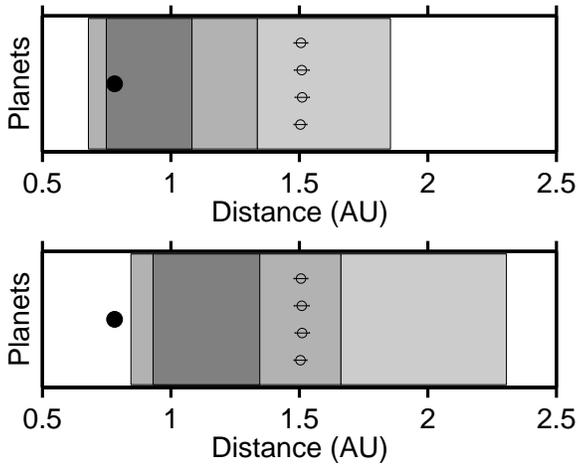}
\end{center}
\caption{
Distances of 55~Cnc~f (full circle) and the hypothetical planet 3
(open circles) in relationship to the HZ of 55~Cnc, where dark gray,
medium gray and light gray denote the conservative, general, and extreme HZ,
respectively (see Table~\ref{HabZ}).  The position of HPL~3
was computed following the TBR based on MDIF~1, 2, 3, and 4
(from top to bottom).  The depicted uncertainty bars are 3$\sigma$, and
are mostly due to the small observational uncertainties of 55~Cnc~f and 55~Cnc~d.
The top figure shows the HZ of 55~Cnc based on $R=0.925~R_{\solar}$ \citep{rib03},
whereas the bottom figure assumes $R=1.15~R_{\solar}$ \citep{bai08}, implying a
relatively high stellar luminosity of 0.90~$L_{\solar}$.
}\label{fig3}
\end{figure}

\begin{table}
\caption{Convergence of Titius--Bode Parameters\label{ConvTBPar}}
\begin{center}
\begin{tabular}{lccc}
\hline
\noalign{\smallskip}
MDIF   & Precision   &   $Z^a$     & Residual \\
\noalign{\smallskip}
\hline
\noalign{\smallskip}
MDIF~1   & $10^{-1}$ &  2.1{\y1}   &  $4.41 \times 10^{-1}$  \\
         & $10^{-2}$ &  1.98{\x1}  &  $3.90 \times 10^{-2}$  \\
         & $10^{-3}$ &  1.973{\p1} &  $3.31 \times 10^{-2}$  \\
         & $10^{-4}$ &  1.9704     &  $3.03 \times 10^{-2}$  \\
MDIF~2   & $10^{-1}$ &  2.0{\y1}   &  $3.93 \times 10^{-1}$  \\
         & $10^{-2}$ &  1.98{\x1}  &  $2.04 \times 10^{-1}$  \\
         & $10^{-3}$ &  1.971{\p1} &  $1.91 \times 10^{-1}$  \\
         & $10^{-4}$ &  1.9704     &  $1.90 \times 10^{-1}$  \\
MDIF~3   & $10^{-1}$ &  2.1{\y1}   &  $6.16 \times 10^{-2}$  \\
         & $10^{-2}$ &  1.98{\x1}  &  $5.94 \times 10^{-4}$  \\
         & $10^{-3}$ &  1.969{\p1} &  $4.56 \times 10^{-4}$  \\
         & $10^{-4}$ &  1.9691     &  $4.52 \times 10^{-4}$  \\
MDIF~4   & $10^{-1}$ &  2.0{\y1}   &  $3.80 \times 10^{-2}$  \\
         & $10^{-2}$ &  1.99{\x1}  &  $1.68 \times 10^{-2}$  \\
         & $10^{-3}$ &  1.991{\p1} &  $1.68 \times 10^{-2}$  \\
         & $10^{-4}$ &  1.9908     &  $1.68 \times 10^{-2}$  \\
\noalign{\smallskip}
\hline
\end{tabular}
\end{center}
\begin{list}{}{}
\item[]
$^a$An equivalent increase in precision is also attained for
the parameters $A$ and $B$.
\end{list}
\end{table}

\subsection{The Habitable Zone of 55~Cnc}

In the following we discuss the possible location of a
newly predicted planet in the HZ of 55~Cnc.
The extent of 55~Cnc's HZ can be calculated following
the formalism by \citet{und03} based on previous work by
\citet{kas93}.  \citet{und03} supplied a polynomial fit
depending on the stellar luminosity and the stellar effective
temperature that allows to calculate the extent of the
conservative and the generalized HZ.  Noting that 55~Cnc
is less luminous than the Sun, it is expected that its HZ
is less extended than the solar HZ, for which the limits of
the generalized HZ have been given as 0.84 and 1.67~AU,
respectively \citep{kas93}.

The luminosity of 55~Cnc has been deduced as
$0.61 \pm 0.04$ $L_{\solar}$ based on the stellar distance and
apparent magnitude; note that this value is fully consistent
with the standard values for the stellar effective temperature
and luminosity (see Table~1).  However, if its stellar radius is
1.15~$\pm$~0.035~$R_{\solar}$, obtained by \citet{bai08} based on
new interferometric measurements, instead of $R=0.925~R_{\solar}$
\citep{rib03}, the revised stellar luminosity is thus given as
0.90 $L_{\solar}$.  This increased luminosity is, however, difficult,
though not impossible, to reconcile with the other parameters of
55~Cnc, such as its apparent magnitude, spectral type and distance.

Based on the standard values for the stellar parameters of
55~Cnc, the limits of the conservative HZ are given
as 0.75 and 1.08~AU, whereas the limits of generalized HZ
are given as 0.68 and 1.34~AU (see Fig.~3 and Table~9).
Surely, larger limits are attained if the higher value for
the stellar luminosity is adopted.  The underlying definition
of habitability is based on the assumption that liquid surface
water is a prerequisite for life, a key concept that is also
the basis of ongoing and future searches for extrasolar habitable
planets (e.g., \cite{cat06,coc09}).  The numerical evaluation
of these limits is based on an Earth-type planet with a
CO$_2$/H$_2$O/N$_2$ atmosphere.

We point out that concerning the outer edge of habitability,
even less conservative limits have been proposed in the meantime
(e.g., \cite{for97,mis00}).  They are based on the assumption of
relatively thick planetary CO$_2$ atmospheres invoking strong
backwarming that is further enhanced by the presence of CO$_2$
crystals and clouds.  These types of limits can be as
large as 2.4~AU in case of the Sun; however, they depend on
distinct properties of the planetary atmosphere,
and are thus subject to ongoing studies and controversies 
(e.g., \cite{hal09}).  Concerning 55~Cnc, the revised outer
limit can be as large as 1.85~AU for the standard stellar
parameters and up to 2.31~AU for the increased value of the
stellar luminosity.  These limits can be considered in the
assessment of the hypothetical planet near 1.51~AU
(see Table~7).  Note that the position of this hypothetical
planet is not significantly affected by the choice of MDIF.
Based on the standard parameters for 55~Cnc, this planet would
be located in the extremely extended HZ.  If the increased value
of the stellar luminosity of 0.90 $L_{\solar}$ is assumed, it
would be located in the generalized HZ.  A depiction of the
different scenarios is given in Fig.~3.

Another important aspect regarding the circumstellar habitability
of 55~Cnc concerns the planet 55~Cnc~f.  It is a hot Neptune
with a mass of about 15\% of Jupiter's mass; therefore, it is
almost certainly unfit to host life.  55~Cnc~f is located
at or near the inner edge of the 55~Cnc's HZ, especially
if the standard value for the 55~Cnc's luminosity is adopted.
However, 55~Cnc~f may be an appropriate object for hosting one
or more habitable moons.  Previous studies on habitable moons
of extra-solar planets have been given by, e.g., \citet{wil97}
and \citet{don10}.


\begin{table}
\caption{Distance of the Hypothetical Planet~3\label{DisHP7}}
\begin{center}
\begin{tabular}{lc}
\hline
\noalign{\smallskip}
MDIF  &  Distance~(AU) \\
\noalign{\smallskip}
\hline
\noalign{\smallskip}
 MDIF~1     &  1.506 $\pm$ 0.027 \\
 MDIF~2     &  1.509 $\pm$ 0.029 \\
 MDIF~3     &  1.511 $\pm$ 0.028 \\
 MDIF~4     &  1.504 $\pm$ 0.025 \\
\noalign{\smallskip}
\hline
\end{tabular}
\end{center}
\end{table}

\begin{table}
\caption{Distance of an Hypothetical Exterior Planet\label{DisHExP}}
\begin{center}
\begin{tabular}{lc}
\hline
\noalign{\smallskip}
MDIF  &  Distance~(AU) \\
\noalign{\smallskip}
\hline
\noalign{\smallskip}
 MDIF~1     &  11.35 $\pm$ 0.45 \\
 MDIF~2     &  11.35 $\pm$ 0.44 \\
 MDIF~3     &  11.33 $\pm$ 0.43 \\
 MDIF~4     &  11.66 $\pm$ 0.39 \\
\noalign{\smallskip}
\hline
\end{tabular}
\end{center}
\end{table}

\section{Consideration of Other Studies}

\subsection{Constraints from Orbital Stability Simulations and Comments on
            Planetary Formation}

Any planet proposed to exist in a star--planet system by the means of the TBR
or otherwise needs to pass the test of orbital stability.  Previously, detailed
simulations of orbital stability for the system of 55~Cnc have been given by, e.g.,
\citet{ray08} taking into account the five known planets of the 55~Cnc system.
\citet{ray08} presented a highly detailed study with particular focus on the
effects of mean motion resonances due to the planets 55~Cnc~f and 55~Cnc~d.
They showed that additional planets could exist between these two known planets,
including hypothetical planets located near 1.51 and 2.95 AU.  Specifically,
if two additional planets are assumed to exist between 55~Cnc~f and 55~Cnc~d,
the domains of stability are given as 1.3--1.6 and 2.2--3.3~AU.  This is an
important finding regarding the hypothetical, potentially habitable planet HPL~3,
proposed at 1.51 AU (see Table~7).

A subsequent study of orbital stability was given by \citet{ji09}. They also
considered the impact of mean motion resonances and identified various unstable
locations.  However, they identified a wide region of orbital stability between
1.0 and 2.3~AU, a potential homestead of habitable terrestrial planets.
These potential planets were also identified to have a relatively low
orbital eccentricity.   Furthermore, the orbital stability simulations
by \citet{ray08} and \citet{ji09} are also lending indirect support to the
principal possibility of an exterior planet, which, if existing, should
be located between 10.9 and 12.2~AU from the star (see Table~8).

Although orbital stability for the hypothetical planet near 1.51~AU appears
to be warranted, it is still important to gauge if such a planet could have
formed in the first place noting that the existence of close-in giant planet(s)
have been viewed to have an adverse effect \citep{arm03}.  Subsequent studies
by Raymond et al. (2005, 2006) have pointed out that the formation and
habitability of terrestrial planets in the presence of close-in giant planets
is indeed possible.

\citet{ray06} gave detailed simulations also encompassing the system of
55~Cnc.  Assuming that the giant planets formed and migrated quickly,
they found that terrestrial planets may be able to form from a second
generation of planetesimals.  For 55~Cnc, objects with masses up to
0.6~$M_\oplus$ and, in some cases, substantial water content are able to
form; these objects are found in orbit in 55~Cnc's HZ.  This type of
result is also consistent with findings from the quantitative numerical
program by \citet{wet96}, who demonstrated that habitable terrestrial
planets can form for a wide range of main-sequence stars, noting that
the number and distance distribution of those planets is relatively
insensitive to stellar mass.  This work has been expanded by \citet{ray06}
and applied to the special case of 55~Cnc.  They found that the majority
of the simulations resulted in the formation of terrestrial planets
positioned in 55~Cnc's HZ, including planets near 1.5~AU from the star.

The existence of additional planets, including habitable planets,
in the gap between 55~Cnc~f and 55~Cnc~d, i.e., between 0.781 and
5.77~AU, also appears to be consistent with the existence of a
protoplanetary disk estimated to have a radial surface density profile
such as $\sigma(r) = r^{-3/2}$ \citep{fis08}.  They also discussed
the overall surface density of the disk, which is estimated to be
considerably higher than in the solar nebula.  The formation of
terrestrial planets\footnote{Following the study by \citet{smi09},
any additional planets in the 55~Cnc system are
expected to have a relatively small mass to warrant the system's
orbital stability.  The low mass of any possible planet can also
readily explain why so far they were able to escape detection.}
in the gap is furthermore indirectly implied by the relatively 
high content of heavy elements as indicated by the enhanced
metallicity of 55~Cnc itself given as
[Fe/H] $= + 0.31 \pm 0.04$ \citep{val05}.

\begin{table}
\caption{Habitable Zone of 55~Cancri\label{HabZ}}
\begin{center}
\begin{tabular}{llc}
\hline
\noalign{\smallskip}
Description & \multicolumn{2}{c}{Distance~(AU)} \\
\noalign{\smallskip}
\hline
\noalign{\smallskip}
  ...       & Standard  &    Alternative \\    
\noalign{\smallskip}
\hline
\noalign{\smallskip}
 HZ-i (general)         &  0.68  &     0.84          \\
 HZ-i (conservative)    &  0.75  &     0.93          \\
 HZ-o (conservative)    &  1.08  &     1.35          \\
 HZ-o (general)         &  1.34  &     1.66          \\
 HZ-o (extreme)         &  1.85  &     2.31          \\
\noalign{\smallskip}
\hline
\end{tabular}
\end{center}
\end{table}

\subsection{Previous Application of the TBR to 55~Cnc}

A previous attempt to predict additional planets in the 55~Cnc system
based on the TBR has been given by \citet{pov08}.  At that time all
currently known planets 55~Cnc~b to 55~Cnc~f were already discovered.
\citet{pov08} assumed a ``law" of the form of
\begin{equation}
a_i^{\rm TB} \ = \ \alpha \cdot e^{\lambda i}
\end{equation}
(in AU) with $\alpha$ and $\lambda$ as free parameters.
Taking into account the planets 55~Cnc~e, 55~Cnc~b, 55~Cnc~c, and
55~Cnc~f, located between 0.038 (originally determined position for
55~Cnc~e) and 0.781~AU, and represented by $i=1$ to 4,
$\alpha$ and $\lambda$ were deduced as 0.0148 and 0.9781, respectively.
This modified TBR is successful to also ``predict" the outermost planet
55~Cnc~d at 5.77~AU, counted as $i=6$, which however was well-known when
\citet{pov08} proposed the new law.

Making 55~Cnc~d also part of the fit results in a small adjustment of the
parameters $\alpha$ and $\lambda$, which then acquire the values of 0.0142 and
0.9975, respectively.  The main aspect of the work by \citet{pov08} is to
predict a new planet for $i=5$ and possibly also an exterior planet for
$i=7$ as well.  The corresponding distances of these hypothetical planets
are given as 2.08 and 15.3~AU; the corresponding orbital periods of
the planets are approximately 3.1 and 62 years, respectively.  None of
these putative planets are expected to be located in the 55~Cnc's HZ.
Moreover, the work by \citet{pov08} assumes the planet 55~Cnc~e to be
located at 0.038~AU rather than 0.01583~AU as identified by \citet{win11};
note that \authorcite{pov08} published their work before the updated
result became available.  Thus, it is unclear if or how the ``law"
by \citet{pov08} could be altered to accommodate this finding.  A
less-than-ideal approach could be to assign $i = 0.11$ (or rounded to
$i=0.1$) to the innermost planet 55~Cnc~e.


\section{Evaluation of Statistical Significance}

An important aspect of this study is to evaluate the statistical
significance of the TBR as deduced.  In this case, we largely follow
the method previously applied by \citet{lyn03} to both the Solar
System and the Uranian satellite system.  In the first step, the
mean square deviation is calculated.  For the TBR
proposed in the present work, we find
\begin{equation}
\chi_{\rm fit}^2 \ = \ \frac{1}{4} \sum_i \bigl[ \log(a_{i} - A) -
                       \bigl( {\log}~B + (i - 2) \cdot {\log}~Z \bigr) \bigr]^2
\end{equation}
where $a_{i}$ constitute the observed distances for the planets
$i=3$, 4, 6, and 9.  For $A$, $B$, and $Z$, we adopt the average
values attained by the sets of runs pertaining to the four different
measures of difference, MDIF (see Table~4).  Furthermore, we pursued
a sufficently large number $N$ of statistical model runs, where for
each model $j$ we obtain
\begin{equation}
\chi^2_j \ = \ \frac{1}{4} \sum_i \bigl[ \log(a_{i} - A) -
             \bigl( {\log}~B + (i - 2 + k y_{i}) \cdot {\log}~Z \bigr) \bigr]^2
\end{equation}
with $y_i$ as a random number in the range $[-1/2,+1/2]$ and $k$ taken
as $k=2/3$ \citep{lyn03}.  For the TBR as proposed, we find
$\chi_{\rm fit} =  8.47 \cdot 10^{-2}$; moreover, the average value of
$\chi_j$ is given as $1.56 \cdot 10^{-1}$ with $N = 10^7$.

Equivalent expressions hold if the method suggested by \citet{lyn03}
is applied to the version of TBR previously given by \citet{pov08}.
In this case, we find
\begin{equation}
\chi_{\rm fit}^2 \ = \ \frac{1}{4} \sum_i \bigl[ \log~a_{i} -
                       \bigl( {\log}~\alpha + i \lambda \bigr) \bigr]^2
\end{equation}
for $i=2$, 3, 4, and 6, and
\begin{equation}
\chi^2_j \ = \ \frac{1}{4} \sum_i \bigl[ \log~a_{i} -
             \bigl( {\log}~\alpha + (i + k y_{i}) \lambda \bigr) \bigr]^2 \ .
\end{equation}
For the TBR by \citet{pov08}, we find $\chi_{\rm fit} =  9.67 \cdot 10^{-2}$
and the average value of $\chi_j$ is given as $2.15 \cdot 10^{-1}$.

Evidently, each individual value of $\chi_j$ can either be $\chi_j \le 
\chi_{\rm fit}$ or $\chi_j > \chi_{\rm fit}$.  The amount of $\chi_j$
values with  $\chi_j \le \chi_{\rm fit}$ gives the probability that
the sequence as calculated by the TBR and the observed sequence can
occur by change.  Based on this approach, it is found that the newly
proposed TBR is significant at a level of 91.8\%, whereas the TBR
proposed by \citet{pov08} is significant at a level of 96.5\%.  Thus,
the analysis of statistical significance of these two different
versions of TBR for 55~Cnc shows that both of them are inherently highly
significant albeit the level of significance for the version previously
given by \citet{pov08} is slightly higher.  Moreover, the TBR attained
by \citet{pov08} is based on two free parameters instead of three
free parameters for the description of the planetary sequence.
However, the newly proposed TBR has the attractive feature that
it is of virtually the same form as that of the Solar System; see
Eq. (1).  Here the parameter $B$ describes the compactness of the
system\footnote{The increased compactness of the planetary system of 55~Cnc
compared to the Solar System is already implied by the known five planets,
which inspired coining the term ``packed planetary systems" hypothesis
(e.g., Raymond et al. 2005).}, which is much higher than for the Solar
System.  Furthermore, the parameter $Z$ describes the relative geometrical
spacing of the system; it is virtually identical to that of
the Solar System, i.e., $Z=2.0$.

Perhaps surprisingly, the TBR appears to be about as applicable to
the 55~Cnc system as to the Solar System gauged by whether or not
the degree of similarity between the predicted and observed distance
sequences can occur by chance (see Introduction for data on the Solar System).
Thus, the application of the TBR to 55~Cnc appears to be justified.
However, considerable care is required to draw final conclusions because
the fitting of five known planets assuming four gaps in the sequence is
much easier than fitting the continuous sequence of nine distances,
i.e., with the mean asteroidal distance included and Pluto omitted;
see \citet{nes04}.
It will be the task of future planetary search missions to inquire
if the planetary system of 55~Cnc follows (1) the TBR of this work, (2)
the TBR by \citet{pov08}, (3) another version of TBR-type sequencing, or
(4) no TBR-type sequencing after all, as found for the vast majority of
the currently known planetary systems.


\section{Summary and Conclusions}

The goal of our study is to provide an application of the TBR to
the 55 Cancri star--planet system.  It is clearly understood that
the TBR does not have the same standing as the various
well-established methods for the search and identification of
extrasolar planets; see, e.g., \citet{jon08} for a detailed
overview.  However, as pointed out in the Introduction, the
TBR appears to be successful (within limits) for a large segment
of the Solar System.  Based on principal considerations, it
would therefore be highly extremely unlikely if there was
no other planetary system for which a TBR-type planet spacing
would be realized, although the number of those systems may
be relatively small.

Thus, the predictions conveyed in this study follow the
hypothesis that the TBR is applicable to the 55~Cnc
planetary system.  This approach entails the prediction
of four planetary positions interior to the outermost
planet 55~Cnc~d, which are given as 0.081, 0.41, 1.51,
and 2.95~AU.  Whether these positions are ``empty", filled
with asteroid / comet belts (like in one instant of the
Solar System, where Ceres is customarily viewed as substitute
for a planet) or occupied by a developed planet is beyond
the scope of this work; in this case additional astrophysical
requirements need to be met.

Pertaining to this latter point, the strongest case for
additional planets around 55~Cnc is given for the possible planets
near 1.51 and 2.95~AU, noting that they would be most compatible
with detailed studies of orbital stability \citep{ray08,ji09}
as well as previous studies about 55~Cnc's proposed protoplanetary
disk \citep{fis08}.  For example, \citet{ray08} argued that
additional planets could exist between 55~Cnc~f and 55~Cnc~d,
including planets located near 1.51 and 2.95 AU in the view
of long-term orbital stability simulations.  Specifically,
if two additional planets are assumed to exist between 55~Cnc~f
and 55~Cnc~d, the domains of stability were derived as 1.3--1.6
and 2.2--3.3~AU.  These two ranges are fully compatible with
the results of the present study.

Moreover, the possible planet near 1.51~AU is of particuar
interest as it appears to be located at the outskirts of
the stellar HZ.  The existence of this planet is consistent
with both orbital stability simulations and studies of the
formation of terrestrial planets for 55~Cnc.  For example,
calculations of terrestrial planet formation by \citet{ray06}
showed that the majority of their simulations lead to the
build-up of terrestrial planets in 55~Cnc's HZ, including
planets near 1.5~AU from the star.  Nonetheless, the final
verdict of potential habitability of this possible planet
(if existing) depends on a sizeable number of factors,
including the definition of the zone of circumstellar
habitability, considering that the outer boundary of
the stellar HZ is notably impacted by a variety of
astrobiological aspects (e.g., \cite{lam09}) including
the structure, density and compositions of the planetary
atmosphere (e.g., \cite{for97,mis00,hal09}).

The TBR as implemented follows closely the previous
version as applied to the Solar System with $A$, $B$,
and $Z$ as free parameters, which were obtained
through careful statistical fitting considering the five
known planets of the 55~Cnc system.  Here we obtain
$A \simeq 0.031$, $B \simeq 0.049$, and $Z \simeq 1.98$.
The parameter $B$ indicates the compactness of the
planetary system, which is much higher than for the
Solar System as also indicated by the conventional version
of Solar System's TBR ($B=0.3$).
The parameter $Z$ indicates the relative geometrical spacing
of the planetary system; it is virtually identical to that
of the Solar System ($Z=2.0$).
We also computed the orbital distance of a the next possible
outer planet of the 55~Cnc system, i.e., exterior to 55~Cnc~d.
Our study predicts that if existing it would be located between
10.9 and 12.2~AU from the star.  This distance is consistent
with previous orbital stability analyses, which predict
the position of a possible exterior planet as 
beyond 10~AU \citep{ray08}.

In summary, it is obviously up to future observational efforts to
verify or falsify the existence of any possible additional planets.
From a general perspective, these efforts will assist in gauging
the applicability of the TBR outside of the Solar System, and will
help to discriminate between different versions of the TBR if found
applicable to the 55~Cnc star--planet system.

\bigskip

The author acknowledges valuable comments by an anonymous referee, which
also resulted in the addition of Section~4.


\end{document}